\documentclass[conference]{IEEEtran}
\IEEEoverridecommandlockouts
\usepackage{cite}
\usepackage{amsmath,amssymb,amsfonts}
\usepackage{algorithmic}
\usepackage{graphicx}
\usepackage{textcomp}
\def\BibTeX{{\rm B\kern-.05em{\sc i\kern-.025em b}\kern-.08em
    T\kern-.1667em\lower.7ex\hbox{E}\kern-.125emX}}
\begin{document}

%\title{BlendSPS: A BLockchain-ENabled Decentralized Security Microservices for Smart Public Safety}

\title{BlendMAS: A BLockchain-ENabled Decentralized Microservices Architecture for Smart Public Safety}

\author{
\IEEEauthorblockN{Ronghua Xu, Seyed Yahya Nikouei, Yu Chen}
\IEEEauthorblockA{%\textit{Dept. of Electrical \& Computer Engineering} \\
\textit{Binghamton University, SUNY} \\ Binghamton, NY 13902, USA \\
\{rxu22, snikoue1, ychen\}@binghamton.edu}
\and
\IEEEauthorblockN{Erik Blasch, Alex Aved}
\IEEEauthorblockA{\textit{US Air Force Research Laboratory} \\
Rome, NY 13441, USA \\
\{erik.blasch.1, alexander.aved\}@us.af.mil}
%\and
%\IEEEauthorblockN{Genshe Chen}
%\IEEEauthorblockA{\textit{Intelligent Fusion Technology, Inc.} \\
%Germantown, MD 20876, USA \\
%gchen@intfusiontech.com}
}

\maketitle

\begin{abstract}
Thanks to rapid technological advances in the Internet of Things (IoT), a smart public safety (SPS) system has become feasible by integrating heterogeneous computing devices to collaboratively provide public protection services. While a service oriented architecture (SOA) has been adopted by IoT and cyber-physical systems (CPS), it is difficult for a monolithic architecture to provide scalable and extensible services for a distributed IoT based SPS system. Furthermore, traditional security solutions rely on a centralized authority, which can be a performance bottleneck or single point failure. Inspired by microservices architecture and blockchain technology, this paper proposes a BLockchain-ENabled Decentralized Microservices Architecture for Smart public safety (BlendMAS). Within a permissioned blockchain network, a microservices based security mechanism is introduced to secure data access control in a SPS system. The functionality of security services are decoupled into separate containerized microservices that are built using a smart contract, and deployed on edge and fog computing nodes. %Implemented and tested on a prototype including edge nodes (Raspberry Pi) and fog nodes (desktop and laptop), 
An extensive experimental study verified that the proposed BlendMAS is able to offer a decentralized, scalable and secured data sharing and access control to distributed IoT based SPS system.
\end{abstract}


%===================================== 1. Introduction ============================================

\section{Introduction}
\label{sec:intro}  % \label{} allows reference to this section
The proliferation of Internet of Things (IoT) technology allows the concept of Smart Cities to become feasible with smart surveillance as one of the most intensively studied topics in the IoT community. Smart Public Safety (SPS) systems process surveillance video streams at the edge and utilize many smart sensors. However, there are still challenges to be tackled in order to realize a fully functional IoT-based SPS system in practice. Relying on a centralized architecture that is based on cloud computing center inevitably adds uncertain latency and poses extra workload to communication networks. While merging lower-level image processing tasks with an edge computing platform is able to meet the requirements for delay-sensitive, mission-critical applications \cite{chen2017enabling}, \cite{nikouei2018lcnn}, new challenges are also introduced by the distributed and cross-domain features such as scalability, heterogeneity and interoperability.

The SPS system is deployed in a distributed network environment that includes a large number of IoT devices (camera + edge hardware) with high heterogeneity and dynamics. The heterogeneity and resource constraint at edge necessitate a scalable, flexible and lightweight system architecture that supports fast development and easy deployment among multiple service providers. Furthermore, those smart devices are geographically scattered across near-site network edges. It is therefore not suitable to enforce security policies on a centralized authority basis, which suffers from the performance bottlenecks or single point of failures. Thus, the SPS system needs a decentralized framework that provides a security mechanism in the trust-less network environments.

Recently, a novel service oriented architecture (SOA), called the microservices architecture, has emerged and gained a lot of popularity \cite{krylovskiy2015designing} in designing a smart city platform. Instead of deploying the system as a monolithic unit as traditional SOAs do, the microservices architecture divides an monolithic application into multiple atomic microservices that run independently on distributed computing platforms. Each microservice performs one specific sub-task or service and requires lightweight communication with other system components. Such characteristics make the microservices architecture an ideal candidate to build a flexible platform, which is easy to be developed and maintained for cross-domain applications. Specifically, the microservices architecture possesses many attractive features, such as good scalability, fine granularity, loose coupling, continuous development, low maintenance cost, and so on. These beneficial features make a microservices architecture a natural selection to enhance SPS systems based on the edge computing paradigm.

A microservices-enabled application also demonstrates vulnerabilities in security due to its usage of distributed data sharing and accessing interfaces \cite{yu2018survey}. Recent methods demonstrate efforts in developing new decentralized security solutions for distributed network applications. Blockchain, which acts as the fundamental protocol of Bitcoin \cite{nakamoto2008bitcoin}, has demonstrated great potential to revolutionize the fundamentals of information technology (IT) due to many attractive properties, such as decentralization and transparency. Decentralized Application (DApp), which is built on smart contract and deployed on blockchain network, performs pre-defined algorithms and agreement without relying on third-party intermediary. Blockchain and smart contract together are promising to provide a decentralized solution to enable a secured data sharing and access control for SPS systems.

In this paper, a BLlockchain-ENable Decentralized Microservices Architecture for SPS (BlendMAS) is proposed to secure data accessing among different service providers and entities in a public safety system. The proposed platform follows the divide-and-conquer principle to decouple the SPS and security functionality into multiple containerized microservices that are computationally affordable to each individual computing platform. The distributed microservices could cooperate with each other as a service pool to perform complicated decision-making and analytical missions. The mining services enforce a consensus mechanism among a large number of authorized miners to maintain sanctity of the data recorded on the permissioned blockchain network. The security mechanism of the SPS is implemented as separated microservices that are built on the smart contract. The hash of the frame features and the corresponding decision is put into a block data, that is approved and appended to the blockchain network for data tampering prevention. The identity authentication and access control strategy ensure that only an authorized entity is capable of accessing services and data in a SPS system.

The major contributions of this paper are:

\begin{itemize}
\item[1)] A complete architecture of microservice-based SPS platform is introduced, which is implemented in a hierarchical edge-fog-cloud computing paradigm;

\item[2)] A fully functional permissioned blockchain network is implemented as microservices and deployed on a physical network; 

\item[3)] A prototype of smart contract enabled data sharing and access control mechanism is designed and tested on a permissioned blockchain network; and

\item[4)] A comprehensive experimental study has been conducted that compares the proposed BlendMAS with the monolithic SOA framework. The experimental results validate the feasibility of the BlendMAS scheme in IoT environments without introducing significant overhead.
\end{itemize}

The remainder of this paper is organized as follows: Section \ref{sec:related} analyzes and reviews the state of the art research and on-going effort in each of components adopted in a SPS system. Section \ref{sec:BlendMAS} illustrates the details of the proposed BlendMAS system. Beside the implementation of the proof-of-concept prototype, Section \ref{sec:experiment} reports an extensive experimental study using test scenarios that are built on both edge devices (Raspberry Pi) and fog computing devices (Desktop). Finally, a summary is presented in Section \ref{sec:conclusion}.

% =========================================== 2.related work =============================================
\section{Background Knowledge and Related Work}
\label{sec:related}  % \label{} allows reference to this section

\subsection{Smart Public Safety}
\label{subsec:sps review}

Traditional surveillance systems depend on human operators to interpret the processing of captured video \cite{chamasemani2013systematic}. However, there is a growing demand for human resources to monitor the data stream as the camera numbers rise in congested areas \cite{blasch2012high}. Recently a number of smart systems are introduced which aim at minimizing the role that human operators play in object detection, such that the responsibility of abnormal behavior detection is taken by various more intelligent machine learning (ML) algorithms~\cite{wang2013intelligent}. Some techniques employ statistical analysis~\cite{fuse2017statistical} or~\cite{ribeiro2017study} that use more modern ML approaches where the algorithm automatically processes the collected video frames in a cloud to detect, track, and report any unusual circumstances.

These traditionally algorithms are computationally expensive and normally implemented at the powerful cloud servers of the surveillance system. An example is the Wide Area Motion Imagery (WAMI) that transforms the frames from the image sensors back to the cloud for processing \cite{chen2016real}, \cite{wu2014container}, \cite{wu2015pseudo}, \cite{wu2017container}. Earlier studies show that this approach puts a heavy burden on the network ~\cite{chen2017enabling}, \cite{chen2016smart}. Ideally, the minimum delay and communication overhead is achievable if all the functions are conducted on-site at the network edge, and the decision is made instantly.

Recently, the smart surveillance community has introduced some decentralized surveillance frameworks that are more convincing in many mission-critical, delay sensitive tasks \cite{nikouei2018eiqis}, \cite{xu2018real}. Implemented based on the edge-fog-cloud hierarchy architecture. the input frame that is streamed out of the surveillance camera is given to an edge unit where low-level processing is performed \cite{nikouei2018kerman}, \cite{nikouei2018intelligent}. The intermediate-level is fog nodes, where multiple tasks are performed based on the processing power and resources available. Finally, the cloud is focused on historical profile building, algorithm fine tuning, and global statistical analysis, which depends on the type of decisions the system is going to make. 

%--------------------2.2 Microservices introduction --------------------
\subsection{Microservices in IoT}

A service oriented architecture (SOA) is widely adopted in the development of application software in a IoT and CPS environment \cite{butzin2016microservices}. The traditional SOA utilizes a monolithic architecture that constitutes different software features in a single interconnected and interdependent application and database. Owing to the tightly coupled dependence among functions and components, such a monolithic framework is difficult to adapt to new requirements in an IoT-enabled system, such as scalability, service extensibility, data privacy, and cross-platform interoperability \cite{datta2018next}. As an extension of the traditional SOA, the \textit{microservices architecture} allows functional units of an application to work independently with a loose coupling though encapsulating a minimal functional software module as a microservice, which can be individually developed and deployed. Each microservice is a process dedicated to certain function of the application. The individual microservices communicate with each other through a lightweight mechanism, such as HTTP RESTful API or a message bus asynchronously \cite{lu2017secure}. Finally, multiple decentralized individual microservices cooperate with each other to perform the functions of complex systems. The flexibility of microservices enables continuous, efficient, and independent deployment of application function units. As two significant features of the microservices architecture, \textit{fine granularity} means each of the microservices can be developed in different frameworks and with minimal development resources, while \textit{loose coupling} implies that functions of microservices components is independent of each other’s deployment and development \cite{yu2018survey}.

Thanks to granularity and coupling properties, the microservices architecture has been investigated in many smart solutions to enhance the scalability and security of IoT-based applications. The IoT systems are advancing from ``things''-oriented ecosystems to a widely and finely distributed microservices-oriented ecosystems \cite{datta2018next}. An Intelligent Transportation Systems (ITS) that incorporates and combines the IoT approaches using the serverless microservices architecture has been designed and implemented to help the transportation planning for the Bus Rapid Transit (BRT) systems \cite{herrera2018smart}. To enable a more scalable and decentralized solution for advanced video stream analysis for large volumes of distributed edge devices, a conceptual design of a robust smart surveillance systems was proposed based on microservices architecture and blockchain technology \cite{nagothu2018microservice}. It aims at offering a scalable, decentralized and fine-grained access control solution for smart surveillance systems. 

%--------------------2.3 Blockchain and Smart contract introduction --------------------
\subsection{Blockchain and Smart Contract}

As a fundamental technology of Bitcoin \cite{nakamoto2008bitcoin}, \textit{blockchain} initially was used to promote a new cryptocurrency that performs commercial transactions among independent entities without relying on a centralized authority, like banks and government agencies. Essentially, the blockchain is a public ledger based on consensus rules to provide a verifiable, append-only chained data structure of transactions. Thanks to the decentralized architecture that does not rely on a centralized authority, blockchain allows the data to be stored and updated distributively. The transactions are approved by miners and recorded in the time-stamped blocks, where each block is identified by a cryptographic hash and chained to preceding blocks in a chronological order. In a blockchain network, a \textit{consensus mechanism} is enforced on a large amount of distributed nodes called miners to maintain the sanctity of the data recorded on the blocks. Thanks to the “trustless” proof mechanism running on miners across the network, users can trust the system of the public ledger stored worldwide on many different decentralized nodes maintained by ''miner-accountants'', as opposed to having to establish and maintain trust with a transaction counter-party or a third-party intermediary \cite{swan2015blockchain}. Thus, blockchain is an ideal decentralized architecture to ensure distributed transactions among all participants in a trustless environment, like edge-based IoT networks.

%While blockchain has shown its success in decentralization of e-currency and payments, extending blockchain's applications beyond the scope of cryptocurrency becomes a trend by designing programmable contracts that support variety of flexible transaction types. 
Emerging from the intelligent property, a \textit{smart contract} allows users to achieve agreements among parties through a blockchain network. By using cryptographic and security mechanisms, a smart contract combines protocols with user interfaces to formalize and secure relationships over computer networks \cite{szabo1997formalizing}. A smart contract includes a collection of pre-defined instructions and data that have been saved at a specific address of blockchain as a Merkle hash tree, which is a constructed bottom-to-up binary tree data structure. Through exposing public functions or application binary interfaces (ABIs), a smart contract interacts with users to offer the predefined business logic or contract agreement. The blockchain and smart contract enabled security mechanism for applications has been a hot topic and some efforts have been reported recently, for example, smart surveillance system \cite{nagothu2018microservice, nikouei2018realtime}, social security system \cite{xu2018constructing}, space situation awareness \cite{xu2018exploration}, identification authentication \cite{hammi2018bubbles} and access control \cite{xu2018blendcac, xu2018smartcac}. Blockchain and smart contract together are promising to provide a solution to enable a secured data sharing and access authorization in decentralized SPS systems.

%------------- 3 -System design of BlendMAS ---------------------
%\section{BlendMAS: a BLockchain-ENabled Decentralized Microservices Architecture for SPS}
\section{System Design of BlendMAS}
\label{sec:BlendMAS}  % \label{} allows reference to this section

%Fig1_System_architecture_BlendMAS
\begin{figure} [t]
\begin{center}
\begin{tabular}{c}
\includegraphics[height=12.4cm]{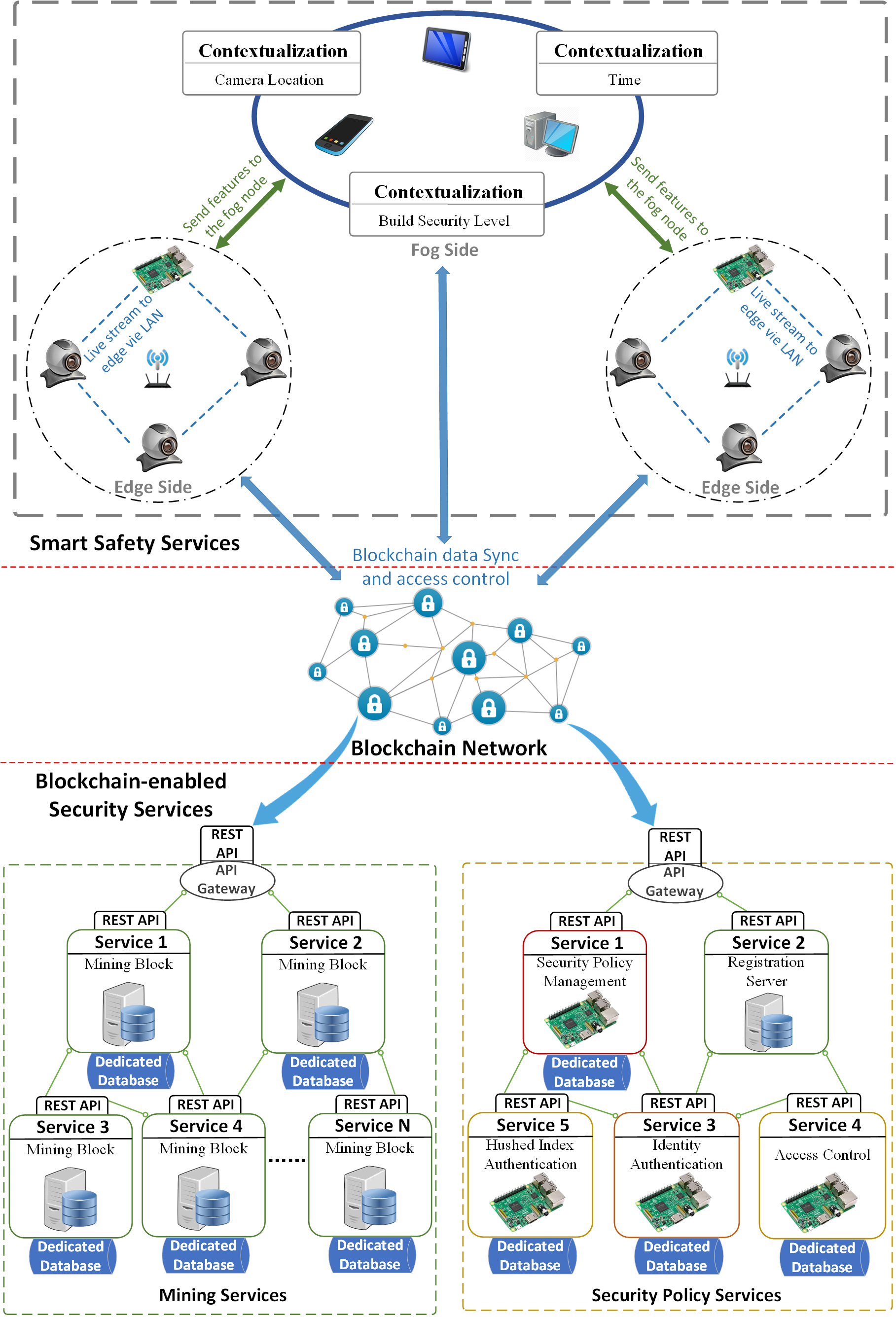}
\end{tabular}
\end{center}
\caption[example] {Illustration of the BlendMAS System Architecture.}
\label{fig:1-BlendMAS} 
\vspace{-10pt}
\end{figure}

Leveraging the attractive characteristics of the microservices architecture that support fine granularity, loose coupling, and continuous delivery, the BlendMAS system is a completely decentralized solution where individual functional components of system are developed by different teams and hosted by heterogeneous hardware platforms, as shown in Fig. \ref{fig:1-BlendMAS}. In the system design, the Docker container is adopted for the microservices architecture and the multi-layer BlendMAS platform is implemented following the edge-fog-cloud computing paradigm. Two type of containers are deployed at the edge layer. One is responsible for the security policy service that enforces the data access control and verification to prevent from unauthorized service request and data tampering, while another is the video stream processing microservice to extract features of frames. Owing to more powerful computing and storage resources, the features fusion, behavior analysis and mining microservices are hosted at the fog layer or cloud layer.

 %------------- 3.1-System architecture of BlendCAC ---------------------
\subsection{System Architecture of BlendMAS}

Figure \ref{fig:1-BlendMAS} illustrates the proposed BlendMAS system architecture, which utilizes microservices-enabled private blockchain network to secure video stream services while providing secured data sharing. The proposed system consists of three services:

\begin{itemize}
   \item \emph{Smart Surveillance Application Services}: These services provide functions to support smart surveillance, such as video stream processing, object detection and tracking, and movement features extraction. Real-time video streams are generated by cameras and transmitted to edge microservices for features extraction. Lower level features are transferred to fog nodes for data aggregation and higher level analytic services, such as pattern recognition, behavior analysis and anomalous event detection.
   
   \item \emph{A Permissioned Blockchain Network}: The security microservice provides not only network communication channel on the Internet, but also a private blockchain network infrastructure running on a decentralized peer-to-peer network. All the network communications among entities run on TCP/IP protocol. Security solutions, such as identity authentication and access control, are developed as Decentralized Application (DApp) which is based on a smart contract and deployed on the blockchain network.
   
   \item \emph{Blockchain-enabled Security Services}: The security service section acts as a fundamental service pool to support key functions of security mechanism. The provided services could be divided into two main clusters: mining services and security policy services. As a core function of maintaining the blockchain network, mining services are responsible for executing consensus algorithms to verify transactions and generating new blocks. The miners are containerized microservices running on single or multiple host machines to fulfill mining task independently. Finally, multiple certificated miners cooperate with each other to secure the private permissioned blockchain network. All the security polices and models, such as identity authentication and access control, are transcoded into separate microservices, and those microservices work together as a security policy services cluster. Through implementing each security model or policy as a single microservice that works independently from each other, the security policy services cluster could address scalability and heterogeneity in IoT-based smart surveillance systems by offering more flexible, interoperable and lightweight security solutions.  
 \end{itemize}

In the SPS system, the feature data extracted by the edge computing will be merged with contextual data on fog layer nodes for high level tasks (such as situation awareness). Therefore a security mechanism is necessary to protect the data shared among functional service nodes and enforce an access control policy. To enable a decentralized, scalable, and fine-grained security scheme for SPS system, the proposed BlendMAS is focused on two issues: the permissioned blockchain management, and security mechanism enforcement.

%------------------------ 3.2-Identity based Permissioned Blockchain ------------------------------------
\subsection{Identity-based Permissioned Blockchain}
All the entities on the permissioned blockchain network are implemented as containers, which perform blockchain services independently on the host machines. The containerized microservices could be categorized as miners or non-mining nodes given the computation power of the host machines. Only the authorized participants could be recognized by entities of network and perform blockchain services, such as mining blocks, sending transactions and deploying smart contracts. An identity management mechanism ensures that participants identify each other while not necessarily fully trust each other.

%Fig2_mining_service
\begin{figure} [t]
\begin{center}
\begin{tabular}{c}
\includegraphics[height=7.2 cm]{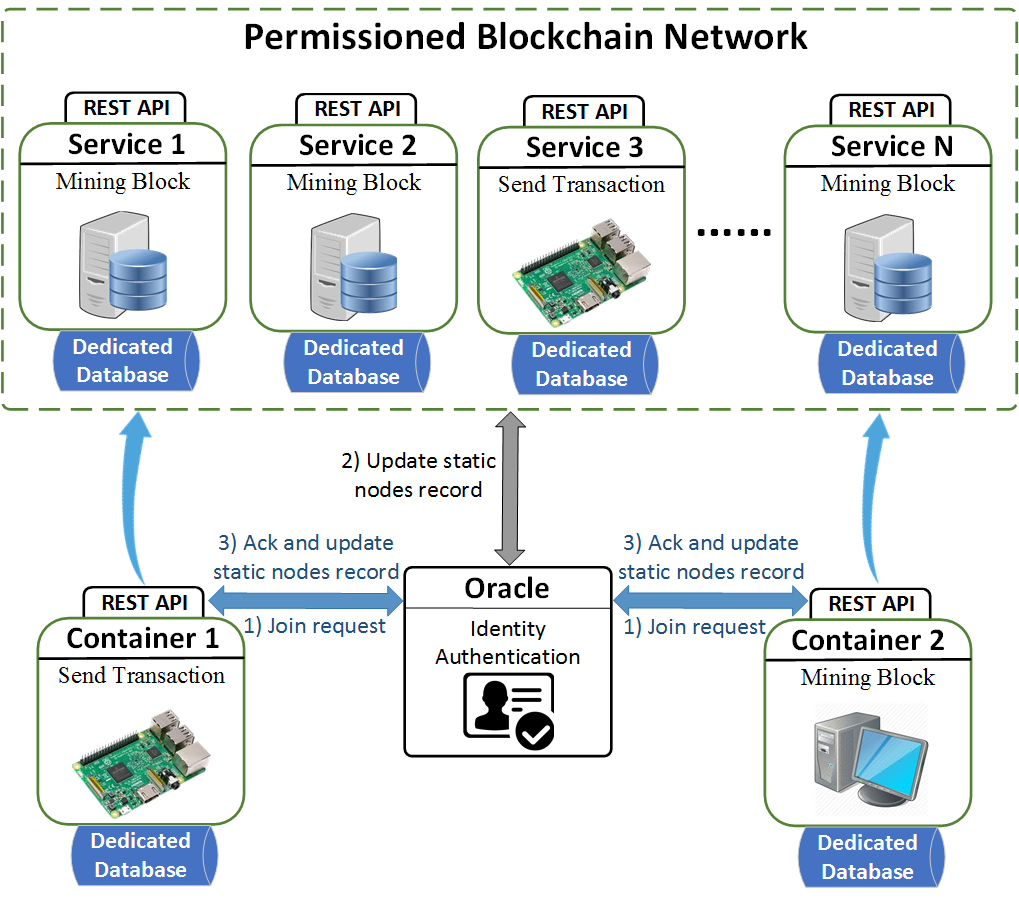}
\end{tabular}
\end{center}
\caption[example] { \label{fig:2-miningsrv} Identity-based Permissioned Blockchain Network.}
\vspace{-10pt}
\end{figure}

Figure \ref{fig:2-miningsrv} illustrates the identity authentication process to enroll new node in the permissioned blockchain network. An oracle, who acts as the administrator of blockchain network, maintains a global node registration and identification policies for the permissioned blockchain network management. To join the permissioned blockchain network, the participants must send joining requests to an oracle for identity authentication. The oracle verifies new entity's joining request by performing identification policies. After the oracle approved the joining requests, the new entity's node information will be added to a global static node record, and the oracle will send updated static nodes record to all certificated participants in blockchain network accordingly. The membership revocation occurs when an entity explicitly launches leaving request or oracle implicitly rule out any misbehaved node. The oracle simply updates static node record on each participant to change configuration of blockchain network. As shown in Fig. \ref{fig:2-miningsrv}, all the containerized miners cooperate with each other to enforce consensus mechanism, while other non-mining containers function as service providers to offer blockchain interactive services like sending transactions. Compared to a public blockchain network, the permissioned blockchain network can achieve a more efficient consensus mechanism, and more secured network by limiting participants with clearly defined security policies.

%------------------------ 3.3-Security Services ------------------------------------
\subsection{Blockchain-enabled Security Microservices}

Utilizing the microservices architecture, the security functions are decoupled into multiple microservices and deployed on distributed computing devices. These decentralized security microservices work as a service cluster to offer a scalable, flexible and lightweight data sharing and access control mechanism for the SPS system. The key service components and operations are introduced below. 

\subsubsection{Registration Service} In the proposed BlendMAS system, all entities must sent registration request to registration microservice before accessing smart surveillance services. Entity registration process is performed by the registration microservices associating entity's unique blockchain account address with a Virtual ID (VID). All the registration information indexed by VID are recorded in a profile database maintained by the registration microservices on the fog node. 

\subsubsection{Identity Authentication} Since each blockchain account is uniquely indexed by its address that is derived from his/her own public key, the account address is ideal for identity authentication needed by other security microservices, such as hashed index recording and access control. The identity authentication microservices expose RESTful API to other microservices-enabled providers for referring identity verification results. Once an authentication service request is received, the identity authentication decision making process queries the requester identity profile from the registration service, and returns the identity verification results according to the pre-define policies.

\subsubsection{Security Management} In the security services cluster, the security management microservices act as data and security service managers who deploy the smart contracts encapsulating hashed index authentication and the access control policies. Taking advantage of the cryptographic and security mechanisms provided by the blockchain network, smart contracts can secure any algorithmically specifiable protocols and relationships from possible malicious interference by third parties in a trustless network. After the smart contracts have been deployed successfully on the blockchain network, they become visible to the entire network. The authorized participants could call Remote Procedure Call (RPC) interfaces to interact with smart contract.

\subsubsection{Hashed Index Authentication} To successfully save the generated hashed index record to the blockchain, a hash index recording microservices entity initially sends an access request to the security management microservices to get a permission for executing the hashed index record generation application binary interfaces (ABI) functions of a smart contract. If the access request is granted, the index recording microservices receive acknowledgement from security management microservices with a smart contract address and the ABI function for recording hashed index data. After generating hashed index records, the hashed index recording microservices simply interact with the authorized ABI function to save the hashed index data to the blockchain. In hashed index authentication process, microservice queries a hashed key-value index by interacting with smart contract and compares it with calculated hash values of the record index table. Finally, the authentication results are sent back to service requester.

\subsubsection{Access Control} To successfully access services or resources at SPS service providers, an entity initially sends an access right request to the access control microservices to get a capability token. Given a reference of the entity’s profile, which is the authenticated identity information maintained by the registration microservices, a decision making policy module running on the access control microservices evaluates the access request by enforcing the authorization policies. If the access request is granted, the access control microservice issues the capability token encoding authorized access right, and then launches a transaction to update the token data in the smart contract. Once the transaction has been approved and recorded in a new block, the access control microservice responds to the requester by providing a smart contract address for the querying token data. Otherwise, the access right request is rejected and a denied acknowledgement is returned. Authorization validation process is triggered when when a smart surveillance service provider receives a service request from users. Given the validation result that demonstrates the access right policies and conditional constraints are satisfied, the service provider grants the access request and offers services to the requester. Otherwise, the service request is denied.

% ============================================== 4.Implementation and Experiment =========================================
\section{Experimental Analysis}
\label{sec:experiment}  % \label{} allows reference to this section

The BlendMAS system is actually the security sub-system of a complete SPS system. Due to the limited space, the implementation and experimental evaluation of the entire SPS prototype are not reported here. But interested readers may find the experimental results of the microservices architecture based video stream processing at the edge in \cite{nikouei2019decentralized}. 

A concept-proof prototype system has been implemented on a real private Ethereum \cite{ehtereum} blockchain network environment. The smart contract development use Solidity \cite{solidity}, which is a contract-oriented, high-level language for implementing smart contracts. In order to evaluate the performance and the overhead of our proposed access control scheme, the security microservices have been implemented using docker container and deployed on both on edge and fog units. The web service application is based on the Flask framework \cite{flask} using Python, and the profiles and policy rules management are developed using an embedded structured query language (SQL) database engine, called SQLite \cite{sqlite}. 

%------------------------------ Environmental Setup -------------------------------------
\subsection{Environmental Setup}

The mining microservices are deployed on a platform with stronger computing power, like a laptop or a desktop. Two miners are deployed on a laptop and other four miners are distributed on four desktops. Table \ref{tab:testbed} describes configurations of nodes used in the experiments. In this prototype, the laptop acts as a cloud computing server, which takes role of oracle to manage blockchain network. All desktops work as fog computing nodes, while a Raspberry PI 3 Model B runs as edge computing node. The SPS functions and security microservices are hosted both on fog and edge computing. All devices use Go-Ethereum \cite{goethereum} as the client application to work on the blockchain network.

\begin{table}[ht]
\caption{Configuration of Experimental Nodes.} 
\label{tab:testbed}
\begin{center}       
\begin{tabular}{|l|p{1.9cm}|p{1.9cm}|p{1.9cm}|} %% this creates two columns
%% |l|l| to left justify each column entry
%% |c|c| to center each column entry
%% use of \rule[]{}{} below opens up each row
\hline
\rule[-1ex]{0pt}{3.5ex} \textbf{Device} & Lenovo P50 & Dell Optiplex 760 & Raspberry Pi 3 Model B \\
\hline
\rule[-1ex]{0pt}{3.5ex} \textbf{CPU} & 2.3 GHz Intel Core i7 (8 cores) & 3 GHz Intel Core TM (2 cores) & quad-core ARM Cortex A53, 1.2GHz \\
\hline
\rule[-1ex]{0pt}{3.5ex} \textbf{Memory} & 16GB DDR3 & 4GB DDR3 & 1GB SDRAM \\
\hline
\rule[-1ex]{0pt}{3.5ex} \textbf{Storage} & 250G SSD+ 500G HHD & 250G HHD & 32GB (microSD card) \\
\hline
\rule[-1ex]{0pt}{3.5ex} \textbf{OS} & Ubuntu 16.04 & Ubuntu 16.04 & Raspbian GNU/Linux 8 (jessie) \\
\hline
\end{tabular}
\end{center}
\end{table}

%------------------------------ Performance Evaluation -------------------------------------
\subsection{Performance Evaluation}

To evaluate the performance of the microservices-based security mechanism, a service access experiment is carried out on a physical network environment which includes 3 Raspberry PIs and 2 desktops. One Raspberry PI works as a client to send service request, while server side is SPS service provider, who has been both hosted on edge (Raspberry PI) and fog (desktop) nodes. A blockchain enabled capability based access control (BlenCAC) scheme \cite{xu2018blendcac} is selected to enforce the access control policies. The hashed index authentication microservice and access control microservice are deployed on two Raspberry PIs separately. To measure the general cost incurred by the BlendMAS scheme both on the edge device processing time and the network communication delay, 50 test runs have been conducted based on the proposed test scenario, where the client sends a data query request to server for an access permission. This test scenario is based on an assumption that the client has been assigned a valid token when it performs the action on server. Therefore, all steps of hashed index authentication and access right validation must be processed on the server side so that the maximum latency value is computed.

\subsubsection{Computational Overhead} 
\begin{figure}[t]
    \centering
        \includegraphics[width=0.48\textwidth]{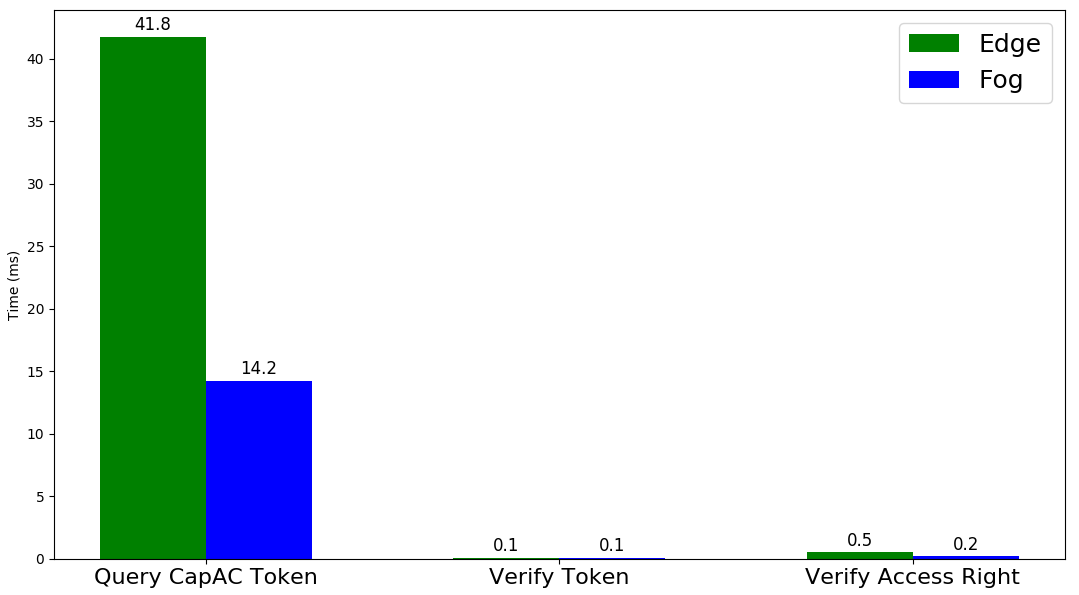}
    \caption{Execution time of each individual stage of the access control Microservices.}
    \label{fig3:CapAC}
%    \vspace{-10pt}
\end{figure}

Figure \ref{fig3:CapAC} shows the computational overhead introduced by AC process. The entire executing time of the access control process is 42.4 ms (41.8 ms + 0.1 ms + 0.5 ms) on the edge device and 14.5 ms (14.2 ms + 0.1 ms + 0.2 ms) on the fog node. The average time for querying data from the smart contract is 41.8 ms on edge device and 14.2 on fog node. Since the authorization process includes CapAC token validation and the access right verification, the average time of authorization process is about 0.6 ms (0.1 ms + 0.5 ms)  on the edge device and 0.3 ms (0.1 ms + 0.2 ms) on the fog node. Owing to cryptography and hash chain computations in data from smart contract, querying CapAC token introduces the highest overload among AC operation stages. The average total delay time cause by service request operation of retrieving data from client to server is 196 ms on the edge device and 85 ms on the fog node. Compared with total delay time, the overload introduced by AC enforcement is about 21 \% (42.4 ms / 196 ms) on the edge side and about 17 \% (14.5 ms / 85 ms) on the fog side.

\begin{figure}[t]
    \centering
        \includegraphics[width=0.48\textwidth]{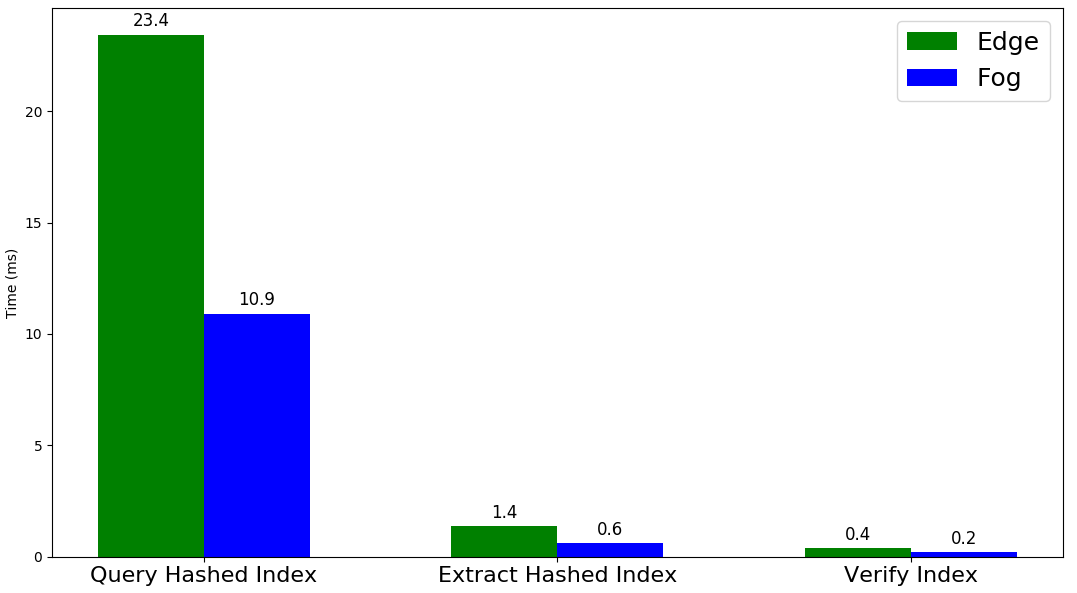}
    \caption{Execution time of each individual stage of the Hashed Index Authentication Microservices.}
    \label{fig4:AuthIndex}
    \vspace{-10pt}
\end{figure}

Figure \ref{fig4:AuthIndex} shows computational overhead incurred by hashed index authentication on both edge and fog sides. Similar to access control process, a querying hashed index operation, which is mainly responsible for fetching hashed index value from smart contract, is the most computing intensive stage. Since the fog nodes have more computation capacity than the edge nodes do, while the execution time of querying hashed index token on edge nodes is about 23.4 ms, the same operation takes only 10.9 ms on fog nodes. The index authentication process is divided into two steps, extracting hashed value from the features data and verifying the hashed index from smart contract given hashed value. The average time of the authorization process is about 1.8 ms (1.4 ms + 0.4 ms) on edge nodes and 0.8 ms (0.6 ms + 0.2 ms) on fog nodes. The average total delay time introduced by the service request operation of retrieving data from client to server is 161 ms on edge device and 51 ms on fog node. Compared with total delay time, the overload introduced by hash index authentication is about 14 \% (23.4 ms / 161 ms) on the edge side and about 21 \% (10.9 ms / 51 ms) on the fog side.

\subsubsection{Communication Overhead} To evaluate the overall network latency incurred by the microservices architecture, a comprehensive test has been performed on two service architectures: the microservices architecture and the monolithic framework. Micro\_App uses microservices framework in which service providers, BlednCAC and hashed index authentication, are decoupled into three separate containers and deployed on different host machines. While Mono\_App uses monolithic framework that all service functions are encapsulated as one container and run on single host machine. Simulating a regular service request allows measuring how long it takes for the client to send a request and retrieve the data from the server. 

\begin{figure}[t]
    \centering
        \includegraphics[width=0.48\textwidth]{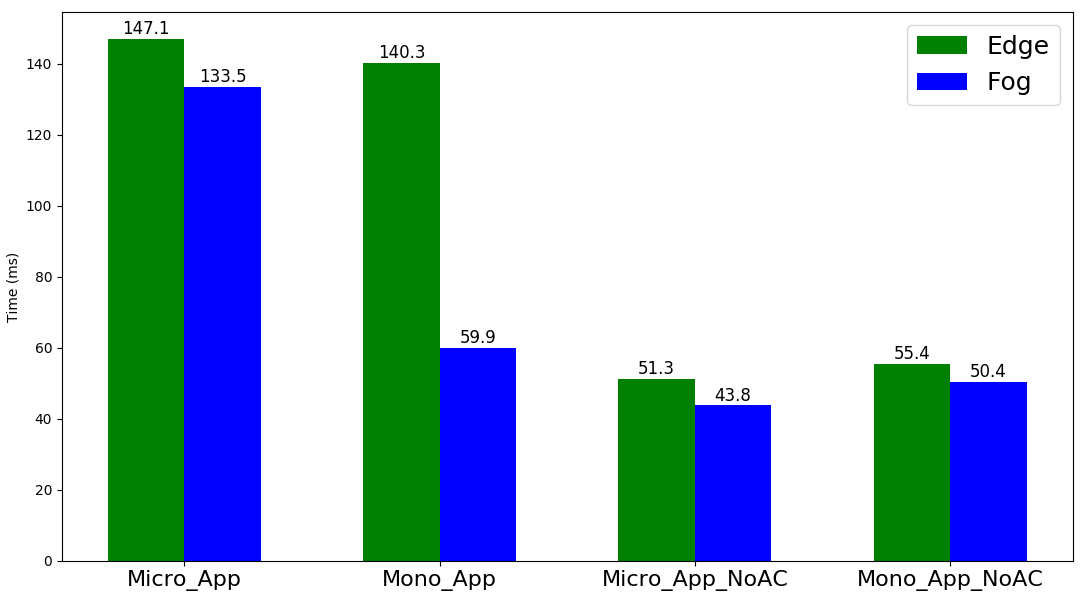}
    \caption{The Network latency incurred by Microservices-enable BlendCAC Solution.}
    \label{fig5:Delay}
    \vspace{-10pt}
\end{figure}

Figure \ref{fig5:Delay} shows the overall communication latency incurred and compares the execution time of the BlendCAC and a benchmark without any access control enforcement on two service architectures. For the monolithic framework, at the fog, the benchmark without access control enforcement (Mono\_App\_NoAC) takes an average of 50.4 ms for fetching requested data versus that the BlendCAC  (Mono\_App) consumes on average of 59.9 ms. It means that embedding the BlendCAC scheme only introduces about 9.5 ms extra latency. At the edge devices, the Mono\_App takes on average of 140.3 ms for fetching requested data versus Mono\_App\_NoAC on average of 55.4 ms, which means that access control process incurs 84.9 ms extra latency. For a microservices architecture, at the fog, the Micro\_App incurs 89.7 ms (133.5 ms - 43.8 ms) extra latency compared with average time consumed by Micro\_App. While at the edge devices, the Micro\_App incurs 95.8 ms (147.1 ms - 51.3 ms) extra latency compared with Micro\_App. No matter which architecture, Mono\_App or Micro\_App, owing to lower computation power of edge computing platform, the network latency incurred by embedding the BlendCAC on edge device is much higher than on fog node. 

Considering the scenarios without access control enforcement, microservices and monolithic framework have almost the same performance both on edge and fog platforms. However, when it comes to BlendCAC scenario, the experiments on two architectures show different communication latency between fog and edge platforms. On the fog computing level, the network latency of Micro\_App takes an average of 133.5 ms for fetching requested data versus that the Mono\_App only consumes on average of 59.9 ms. But for edge devices, both frameworks have almost the same network latency, which is about 140 ms. Since services on monolithic framework utilize an Internal Process Communication (IPC) to call functions and share data, given the same computation and memory power, it is much more reliable and faster than the Remote Process Communication (RPC) used by microservices architecture, which is subject to the Quality of Service (QoS) determined by the network communication status. Since the fog nodes are more powerful than edge devices, the influences caused by computation and network communication for Micro\_App is more significant than the same case for Mono\_App on fog nodes. Although microservices incurs more network latency, which is 73.6 ms (133.5 ms - 59.9 ms) on the fog side and 6.8 ms (147.1 ms - 140.3 ms) on edge side, it still brings benefits to the distributed IoT-based system, such as loosely coupled dependence, easy service deployment and cross-platform interoperability.

\subsection{Discussions}

The experimental results demonstrate that the proposed BlendMAS strategy is effective and efficient to provide secured data sharing and access control services for IoT-based SPS system. Compared to centralized security solutions based on monolithic framework, the BlendMAS scheme has the following advantages:
\begin{enumerate}

\item \emph{Decentralized security mechanism}: leveraging the blockchain and smart contract technology, the proposed BlendMAS scheme allows service providers to control their devices and resource without relying on a centralized third authority to establish the trust relationship with unknown nodes; the risk of performance bottleneck and the single point of failure incurred by centralized architecture are mitigated; and

\item \emph{Fine-grained architecture}: embracing fine-grained microservices architecture enables the smart public safety toward domain-driven design, technology heterogeneity and continuous delivery, which is critical to the scalable, heterogeneous, flexible and dynamic IoT-based smart surveillance applications.

\end{enumerate}

\section{Conclusions}
\label{sec:conclusion}  % \label{} allows reference to this section

In this paper, we proposed a decentralized data sharing and access control mechanism leveraging the microservices and blockchain technology, called BlendMAS, to handle the challenges in IoT-based public safety system. A concept-proof prototype has been built in a physical IoT network environment to verify the feasibility of the proposed BlendMAS. The hash index authentication and access control models are transcoded to smart contracts deployed on the private Ethereum blockchain network. The functionality of smart surveillance and security services are decoupling into separate containerized microservices and deployed on distributed edge and fog computing nodes. Extensive experimental studies have been conducted and the results are encouraging. It validated that the BlendMAS solution is able to efficiently and effectively enforce identity authentication and access control policy in a distributed IoT-based smart surveillance network. This work has demonstrated that the proposed BlendMAS framework is a promising approach to provide a decentralized, scalable, fine-grained architectural security mechanism for SPS system.

While the reported work has shown great potential, there is still a long way to go towards a complete decentralized and lightweight security solution for IoTs and edge computing. Deeper insights are in need. Part of our on-going effort is focused on further exploration of the micro-blockchain platform based on permissioned blockchain network and lightweight consensus algorithm running on IoT devices.

\section*{ACKNOWLEDGEMENTS}

The views and conclusions contained herein are those of the authors and should not be interpreted as necessarily representing the official policies or endorsements, either expressed or implied, of the United States Air Force. 

% References
\bibliographystyle{IEEEtranS} % makes bibtex use IEEEtran.bst
\bibliography{report} % bibliography data in report.bib

\end{document}